\documentclass[
reprint,
superscriptaddress,
amsmath,amssymb,
aps,
]{revtex4-1}

\usepackage{graphicx}
\usepackage{dcolumn}
\usepackage{bm}
\usepackage{todonotes}

\usepackage{color}

\newcommand{\pdiff}[2]{\frac{\partial #1}{\partial #2}}
\newcommand{\fdiff}[2]{\frac{\delta #1}{\delta #2}}

\newcommand{\new}{\nonumber\\}
\newcommand{\abs}[1]{\left|#1\right|}

\newcommand{\hh}{h}
\newcommand{\hr}{\hat{r}}

\newcommand{\bx}{\bm{x}}
\newcommand{\bX}{\bm{X}}
\newcommand{\by}{\bm{y}}

\newcommand{\ave}[1]{\left\langle #1 \right\rangle}

\newcommand{\HH}{\mathcal{H}}

\newcommand{\gap}{\tilde{h}}

\usepackage{amsmath}	
\begin{document}

\preprint{APS/123-Qed} \title{
Jamming and replica symmetry breaking of weakly disordered crystals
}


\author{Harukuni Ikeda}
 \email{hikeda@g.ecc.u-tokyo.ac.jp}
\affiliation{ 
Graduate School of Arts and Sciences, The University of Tokyo 153-8902, Japan
} 


\date{\today}
	     
\begin{abstract}
We discuss the physics of crystals with small polydispersity near the
jamming transition point. For this purpose, we introduce an effective
single-particle model taking into account the nearest neighbor structure
of crystals. The model can be solved analytically by using the replica
method in the limit of large dimensions.  In the absence of
polydispersity, the replica symmetric solution is stable until the
jamming transition point, which leads to the standard scaling of perfect
crystals. On the contrary, for finite polydispersity, the model
undergoes the full replica symmetry breaking (RSB) transition before the
jamming transition point. In the RSB phase, the model exhibits the same
scaling as amorphous solids near the jamming transition point. These
results are fully consistent with the recent numerical simulations of
crystals with polydispersity. The simplicity of the model also allows us
to derive the scaling behavior of the vibrational density of states that
can be tested in future experiments and numerical simulations.
\end{abstract}


\maketitle

\section{Introduction}
Physics of crystal and amorphous solids are qualitatively different. For
instance, low frequency eigenmodes of crystals are phonon, and thus the
vibrational density of states $D(\omega)$ follows the Debye law
$D(\omega)\sim \omega^{d-1}$ where $d$ denotes the spatial
dimensions~\cite{kittel1976introduction}. On the contrary, amorphous
solids have excess non-phonon excitations. As a consequence, the density
of states normalized by the Debye's prediction $D(\omega)/\omega^{d-1}$
shows a peak at a certain frequency $\omega=\omega_{\rm
BP}$~\cite{buchenau1984,malinovsky1986,grigera2003phonon,shintani2008universal,kaya2010}. This
phenomenon is known as the {\it boson peak} and thought to be one of the
universal properties of amorphous solids~\cite{phillips1981amorphous}.

Crystal and amorphous solids also show distinct elastic properties near
the (un) jamming transition point at which constituent particles lose
contact, and simultaneously the pressure vanishes~\cite{van2009jamming}.
Here we focus on the jamming of spherical and frictionless particles
interacting with finite and repulsive potentials. The scaling of these
models is now well understood due to extensive numerical
simulations~\cite{ohern2003,van2009jamming} and
theories~\cite{wyart2005,wyart2005rigidity,charbonneau2014fractal,charbonneau2017glass}.
The shear modulus $G$ of crystals does not show the strong pressure $p$
dependence and remains a constant at the jamming transition
point~\cite{goodrich2014solids,tong2015crystals}. On the contrary, $G$
of amorphous solids shows the power law behavior $G\sim p^{1/2}$ and
vanishes at the jamming transition
point~\cite{ohern2003,van2009jamming}. The behavior of $G$ is directly
related to the contact number per particle $Z$ as $G\propto \delta
Z\equiv Z-Z_{\rm iso}$~\cite{wyart2005rigidity}. Here $Z_{\rm iso}$
denotes the contact number when a system is isostatic, {\it i.e.}, the
number of constraints is the same as the number of degrees of
freedom~\cite{maxwell1864,alexander1998}. At the jamming transition
point, $\delta Z > 0$ for perfect crystals, leading to $G>0$, whereas
$\delta Z = 0$ for amorphous solids, leading to
$G=0$~\cite{van2009jamming,ohern2003}.

Crystal and amorphous are two extreme states of solids: the former is a
state free from disorder while the latter is a state of maximum
disorder. From both theoretical and practical points of views, it is
important to understand how the physical properties shift from that of
crystal to amorphous on the increase of the strength of disorder.
Previous numerical simulations show that small disorder only play a
moderate role far from the jamming transition point ($p\sim 1$). For
instance, numerical studies of crystals with polydispersity show that
the amplitude of the boson peak $D(\omega_{\rm BP})/\omega_{\rm
BP}^{d-1}$ only continuously increases on the increase of the
polydispersity $\eta$, if $\eta$ is small
enough~\cite{mizuno2013elastic,guo2014effect}. Near the jamming
transition point $p\ll 1$, on the contrary, even small disorder
dramatically change the physical properties of crystals. More and more
non-phonon modes appear as $p$ decreases, eventually leading to the
divergence of $D(\omega_{\rm BP})/\omega_{\rm BP}^{d-1}$ in the jamming
limit $p\to 0$, in sharp contrast to perfect crystals where $D(\omega)$
does not show the strong $p$ dependence. Furthermore, for crystals with
small defects or polydispersity, $G$ and $\delta Z$ exhibit the same
power laws of amorphous solids sufficiently near the jamming
transition point~\cite{goodrich2014solids,tong2015crystals}. In
particular, $G$ and $\delta Z$ vanish at the jamming transition point if
there is even infinitesimally small
polydispersity~\cite{mari2009,tong2015crystals}, while for perfect
crystals, $G$ and $\delta Z$ remain finite.

Our aim here is to construct a solvable mean field model being able to
describe the above striking effects of disorder on crystals near the
jamming transition point. We consider a model in the limit of large
dimensions, which is a popular mean field limit in theoretical
physics~\cite{georges1996,parisi2010mean}. In this limit, only the first
virial corrections give a relevant
contribution~\cite{frisch1999,parisi2010mean}. For short-range
potential such as hard spheres, this implies that the information of
nearest neighbor structures is enough to describe the physics. Motivated
by this consideration, we introduce an effective single-particle model
that only takes into account the interactions between a particle of
interest and nearest-neighbor particles. For zero polydispersity
$\eta=0$, our model correctly reproduces the scaling of perfect
crystals. For finite $\eta$, on the contrary, our model predicts that
the existence of the replica symmetry breaking (RSB)
transition~\cite{mezard1987spin} at finite pressure $p=p_{\rm RSB}$. For
$p\leq p_{\rm RSB}$, the model exhibits the same scaling as amorphous
solids. Thereby, our model can reproduce the sharp cross-over from the
scaling of crystal to amorphous observed in previous numerical
simulations of weakly disordered crystals.

\section{Model}
\label{173214_15Jul20}
\begin{figure}[t]
\begin{center}
 \includegraphics[width=6cm]{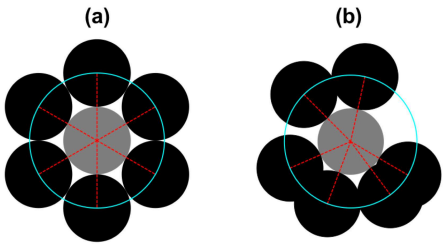} \caption{ Jaming configurations
 for $d=2$ and $\eta=0$.  The gray disk denotes the tracer particle, and
 the black disks denote nearest neighbors. The blue solid line denotes
 the equidistant line from the tracer particle, and the red dashed lines
 denote contacts. (a) Configuration of the hexagonal close-packing. The
 nearest neighbors are arranged periodically on the equidistant
 line. (b) Configuration of our model. The nearest neighbors are placed
 randomly on the equidistant line.}  \label{164607_22Oct19}
\end{center}
  \end{figure}

We consider a tracer particle surrounded by $M$ frozen nearest-neighbor
(NN) particles in the limit of the large spatial dimensions. In $d$
spatial dimensions, the tracer particle has $d$ degrees of freedom,
implying that the model becomes isostatic when the contact number is
$Z=Z_{\rm iso}=d$~\footnote{This is a slightly different condition from
particle systems in $d$ spatial dimensions where all particles are
mobile, and the isostatic condition leads to $Z_{\rm
iso}=2d$~\cite{van2009jamming}.}. The tracer and NN particles interact
with the one-sided harmonic potential:
\begin{align}
 V(\bX) = \sum_{\mu=1}^M v(h^\mu),\
 h^\mu &= \sqrt{d}\left(\abs{\bX-\by^\mu}-\sigma^\mu\right), \label{152502_23Jul18}
\end{align}
where $v(x)=x^2\theta(-x)/2$, and the pre-factor $\sqrt{d}$ of the gap
function $h^\mu$ is necessary to keep $h^\mu = O(1)$ in the $d\to\infty$
limit, see Appendix~\ref{203409_15Jul20} for details. $\bX=\{X_1,\cdots, X_d\}$ and
$\by^\mu=\{y_1^\mu,\cdots,y_d^\mu\}$ denote the positions of the tracer
and $\mu$-th NN particle, respectively.  $\sigma^\mu$ denotes the
interaction range between the tracer and $\mu$-th NN particle.  We
assume that $\sigma^\mu$ can be written as
\begin{align}
 \sigma^\mu = \sigma\left(1+\frac{1}{d}b^\mu\right),\label{182545_8May20}
\end{align}
where $\sigma$ controls the mean size of particles, and $b^\mu$
denotes the polydispersity. The pre-factor of $b^\mu$, $1/d$, is
necessary to keep the relative interaction volume remains finite in the
large dimensional limit: $\lim_{d\to\infty}(\sigma^\mu/\sigma^\nu)^d =
e^{b^\mu-b^\nu}$. $b^\mu$ follows the normal distribution of
zero mean and variance $\eta^2$:
\begin{align}
 P(b^\mu) = \frac{1}{\sqrt{2\pi\eta^2}}\exp\left[-\frac{(b^\mu)^2}{2\eta^2}\right].\label{183903_12May20}
\end{align}
The NN particles are homogeneously distributed on the
surface of the hypersphere of the radius $\sqrt{d}$, namely, the
distribution function of $\by^\mu$ is given by
\begin{align}
 P(\by^\mu) =
 \frac{\delta(\by^\mu\cdot\by^\mu-d)}{\int d\by^\mu\delta(\by^\mu\cdot\by^\mu-d)}.\label{114419_31Mar20}
\end{align}

To get the physical intuition about the model, we first explain the
behavior at the jamming transition point in $d=2$ in the absence of the
polydispersity $\eta=0$, comparing it with the hexagonal packing.  For
the the hexagonal packing, the NN particles are arranged periodically on
the equidistant line from the tracer particle
(Fig.~\ref{164607_22Oct19}(a)).  On the contrary, for our model at the
jamming transition point, the NN particles are randomly distributed on
the equidistant line (Fig.~\ref{164607_22Oct19}(b)). The tracer is in
contact with all NN particles, as in the case of hexagonal packing,
leading to a \textit{hyperstatic} configuration when the number of the NN
particles $M$ is larger than $Z_{\rm iso}$.

The same story holds in general $d$ as long as $\eta=0$: the jamming
occurs when $\sigma\equiv \sigma_J^0=\sqrt{d}$, at which $\bX=0$ and
$h^\mu=0$ for all $\mu$, meaning that the tracer particle is in
contact with all NN particles, leading to a hyperstatic configuration
when $M>Z_{\rm iso}$. On the contrary, for $\eta>0$, the jamming
configuration is non-trivial, which we shall discuss in this manuscript.

\section{Marginal stability}
\label{173234_15Jul20}

The previous works for the mean field models of the jamming transition
unveiled that the systems undergo replica symmetry breaking (RSB) before
reaching the jamming transition point. In the RSB phase the systems are
marginally stable~\cite{mezard1987spin,franz2015universal}. Here we show
that the contact number in the RSB phase can be calculated by using the
marginal stability.

At zero temperature, the stability of the system can be discussed by
observing the Hessian of the interaction potential:
\begin{align}
 \HH_{ij} &= \frac{1}{d}\pdiff{V(\bX)}{X_i\partial X_j}
 \sim \frac{1}{d}\sum_{\mu=1}^M \left(y_i^\mu y_j^\mu 
 + \delta_{ij}h_\mu\right)\theta(-h_\mu),
\end{align}
where the $1/d$ prefactor is necessary to make $H_{ij}=O(1)$, and we
only keep the relevant terms in the limit of $d\to\infty$. In this
limit, $y_i^\mu$ can be identified with the i.i.d. Gaussian random
variable of zero mean and unit variance, see Appendix~\ref{173040_17Jul20}. Thus, $\HH_{ij}$ can
be considered as a Wishart matrix~\cite{livan2018introduction} with an
additional diagonal term. The eigenvalue distribution of $\HH_{ij}$
follows the Marchenko-Pastur distribution~\cite{franz2017universality}:
\begin{align}
 \rho(\lambda) = \frac{1}{2\pi}\frac{\sqrt{(\lambda-\lambda_-)(\lambda_+-\lambda)}}{\lambda+p},\
\lambda_{\pm} =
 \left(\sqrt{z}\pm 1\right)^2 -p,  \label{190507_1Aug18}
\end{align}
where we have defined the contact number per degree of freedom $z=Z/d$
and pressure $p$ as
\begin{align}
 z = \frac{1}{d}\sum_{\mu=1}^M \theta(-h^\mu),\
 p =  -\frac{1}{d}\sum_{\mu=1}^M \theta(-h^\mu)h^\mu.\label{183723_14May20}
\end{align}
In the RSB phase, the marginal stability requires
$\lambda_{-}=0$~\cite{mezard1987spin,franz2015universal}. This
condition with Eq.~(\ref{190507_1Aug18}) determines $z$ as a
function of $p$:
\begin{align}
 z = (1+p^{1/2})^2.\label{144121_25Oct19}
\end{align}
This result implies that (i) the model is isostatic $z=z_{\rm iso}=1$
at the jamming transition point $p=0$~\footnote{The condition of the
isostaticity is now $z=z_{\rm iso}\equiv Z_{\rm iso}/d=1$.}, and (ii)
$z$ exhibits the square root scaling $\delta z = z-z_{\rm iso}\propto
p^{1/2}$ for $p\ll 1$. Those properties are the same as amorphous solids
consisting of soft harmonic particles~\cite{ohern2003} and a mean-field
model of the jamming transition~\cite{franz2017universality}.

Here we used a rather heuristic argument to calculate $z$ in the RSB
phase.  But the same result can be derived by directly solving the RSB
equation, see Appendix~\ref{084946_15Jul20}.

\section{Replica method}
\label{173626_15Jul20}

The RSB is a consequence of the complex structure of the free energy
landscape of amorphous solids near the jamming transition
point~\cite{charbonneau2017glass}. On the contrary, perfect crystals or
nearly perfect crystals have a unique minimum, in other words, the
systems are in the replica symmetric (RS) phase. Since the RS phase is
not marginally stable, we can not use Eq.~(\ref{183723_14May20}).
Instead, we calculate $z$ by using the replica
method~\cite{mezard1987spin}. The calculation is very similar to that of
the perceptron, which was previously investigated as a mean-field model
of the jamming
transition~\cite{franz2016simplest,franz2017universality}. Therefore,
below we just briefly sketch how to calculate $z$ as a function of
$p$. The details of the calculations are provided in
Appendix~\ref{173250_17Jul20}.

To calculate $z$ and $p$, it is convenient to
introduce the gap distribution:
\begin{align}
 g(h) &= \frac{1}{d}\ave{\sum_{\mu=1}^M\delta(h-h^\mu)} = \fdiff{F}{v(h)},
\end{align}
where $F$ denotes the free energy, and $\langle \bullet\rangle$ denotes
the average for both quenched disorder and thermal fluctuation.
Using $g(h)$, $z$ and $p$ are calculated as 
\begin{align}
 z = \int_{-\infty}^\infty dh g(h)\theta(-h),\ 
 p =  -\int_{-\infty}^\infty dh g(h)\theta(-h)h.
\end{align}
We calculate $F$ by using the replica method:
\begin{align}
 -\beta F = \lim_{d\to\infty}\frac{\left[\log \mathcal{Z}\right]_{b,\by}}{d}
 =  \lim_{n\to 0, d\to\infty}\frac{\log\left[\mathcal{Z}^n\right]_{b,\by}}{nd},\label{065054_13May20}
\end{align}
where $\mathcal{Z}$ denotes the partition function:
\begin{align}
 \mathcal{Z} = \int d\bx e^{-\beta V(\bX)}.
\end{align}
$\beta$ represents the inverse temperature. In this work, we investigate
the model only in the zero temperature limit
$\beta\to\infty$. Investigations at finite $\beta$ are left for future
work. The square brackets in Eq.~(\ref{065054_13May20}) denote the
average for the quenched randomnesses:
\begin{align}
 \left[\bullet\right]_{b,\by} =
 \prod_{\mu=1}^M\int db^\mu P(b^\mu)\int d\by^\mu P(\by^\mu)  \bullet,
\end{align}
where $P(b^\mu)$ and $P(\by^\mu)$ are given by
Eq.~(\ref{183903_12May20}) and (\ref{114419_31Mar20}), respectively. By
using the saddle point method, one can represent $F$ as a function of
the correlation among the replicas (see Appendix~\ref{173040_17Jul20}
for details):
\begin{align}
 Q_{ab} = \ave{\bX^a\cdot\bX^b},\ a,b=1,\cdots, n,
\end{align}
where $\bX^a$ and $\bX^b$ denote the positions of the $a$-th and $b$-th
replicas, respectively. As the free energy has a single minimum in the
RS phase, there is no reason to distinguish a specific pair of replicas
$ab$, implying that $Q_{ab}$ is written as
\begin{align}
 Q_{ab} =
 \begin{cases}
  q_0 & (a=b),\\
  q & (a\neq b).
 \end{cases}
 \label{183424_14May20}
\end{align}
We can calculate $q$ and $q_0$ by solving the saddle point equations:
$\partial_{q_0}F = 0$ and $\partial_q F = 0$ (see
Appendix~\ref{173250_17Jul20} for details).  Substituting back the
results to $F$, we obtain the free energy at the saddle point, which
allows us to calculate $g(h)$, $z$, and $p$. Below, we will show the
results only for $M=10d$, but we confirmed that the qualitatively same
results are obtained for different values of $M$.

To see the stability of the RS Ansatz, we calculate the minimal
eigenvalue $\lambda_-$ by substituting the RS results for $z$ and $p$
into Eq.~(\ref{190507_1Aug18}). The RS-RSB transition point is determined
by the condition $\lambda_-=0$.
\begin{figure}[t]
\begin{center}
\includegraphics[width=7cm]{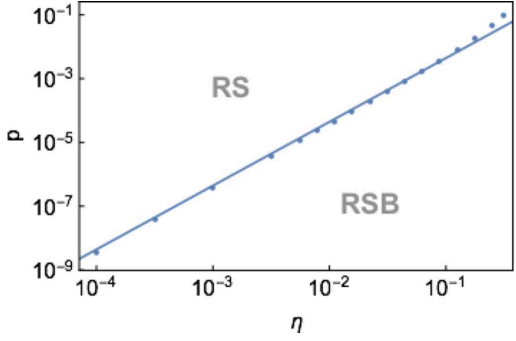} \caption{Phase diagram for
$M=10d$. Markers denote the numerical results, while
solid line denotes a quadratic fit $p\propto \eta^2$.}
\label{110241_25Oct19}
\end{center}
  \end{figure}
In Fig.~\ref{110241_25Oct19}, we show the RS-RSB phase diagram in the
$\eta-p$ plane. It is noteworthy that the RSB always occurs at a finite
pressure $p=p_{\rm RSB}\sim \eta^2$ before reaching the jamming
transition point $p=0$ whenever $\eta>0$.

\section{Scaling of contact number}

\begin{figure}[t]
\begin{center}
\includegraphics[width=9cm]{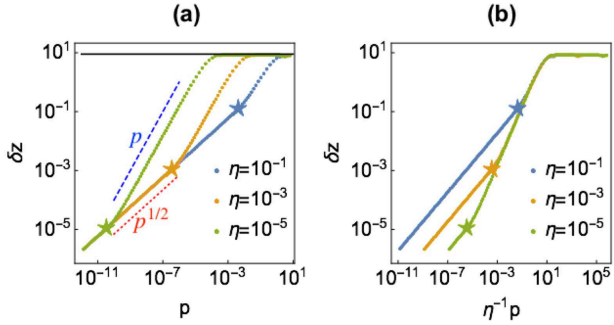} \caption{Scaling of the excess
 contact number $\delta z$ as a function of the pressure $p$ for
 $M=10d$. (a) Filled circles denote the exact results. $\star$ denotes
 the RSB transition point. The black solid line denotes $z=M/d$, the
 blue dotted line denotes $\delta z \sim p$, and the red dashed line
 denotes $\delta z\sim p^{1/2}$.  (b) The same data with the rescaled
 pressure.  } \label{173053_24Oct19}
\end{center}
  \end{figure}  
Following the above procedures, we calculate $z$ for several $\eta$.  We
summarize our results in Fig.~\ref{173053_24Oct19}. There are three
different scaling regions.  For $p\gg \eta$, $z$ takes a constant value
$z\approx M/d$, meaning that the tracer particle contact with most NN
particles, see the black line. For $\eta^2 \ll p \ll \eta$, the contact
number decreases as $\delta z\sim p$, see the blue dotted line.  At
$p=p_{\rm RSB}\sim \eta^2$, the RS solution becomes unstable, and for
$p\leq p_{\rm RSB}$, one should use the RSB result
Eq.~(\ref{144121_25Oct19}). For $p\ll \eta^2 $, 
Eq.~(\ref{144121_25Oct19}) predicts $\delta z\sim p^{1/2}$, see the red
dashed line.

For $p\geq p_{\rm RSB}$, the results for different $\eta$ collapse
on a single curve if one plots $\delta z$ as a function of $\eta^{-1}p$,
see Fig.~\ref{173053_24Oct19} (b). This scaling is consistent with a
previous numerical simulation~\cite{tong2015crystals} and perturbation
theory~\cite{acharya2019athermal}. Remarkably, the above scaling implies
that the two limits $\eta\to 0$ and $p\to 0$ are not commutative: if one
takes the limit $\eta\to 0$ first and then takes the limit $p\to 0$, one
gets $\delta z > 0$, contrary, if one takes the limits in reverse order,
one gets $\delta z = 0$.

\section{Density of states}
\label{174119_15Jul20}
An important quantity to characterize the physics of solids is the
vibrational density of states $D(\omega)$,
which is a distribution of the eigen-frequency $\omega=\sqrt{\lambda}$.
By using Eq.~(\ref{190507_1Aug18}),
$D(\omega)$ is calculated as 
$D(\omega)=2\omega \rho(\omega^2)$. Near the jamming transition point
for small $\omega$, $D(\omega)$ asymptotically behaves as
\begin{align}
 D(\omega) \sim 
 \begin{cases}
  {\rm constant} & \delta z \ll \omega  \ll 1 \\
  \delta z^{-2}\omega^2 & \omega_0< \omega \ll \delta z \\
  0 &                    \omega\leq \omega_0,
 \end{cases}\label{122658_26Oct19}
\end{align}
where $\omega_0 = \sqrt{\lambda_{-}}$. In the RS phase, $\omega_{0}>0$
and $D(\omega)$ has a finite gap~\footnote{In finite $d$, $D(\omega)$
for $\omega<\omega_0$ is described by the Debye theory $D(\omega)\sim
\omega^{d-1}$.}. $\omega_0$ decreases on the decreasing of $p$ and
eventually vanishes at $p=p_{\rm RSB}$. In the RSB phase $p\leq p_{\rm
RSB}$, $\omega_0=0$ and $D(\omega)$ is gapless. For $\omega\ll 1$, the
density of states exhibits the quadratic scaling $D(\omega)\sim
\omega^2$. This is the same result as previous mean field theories of
amorphous solids~\cite{degiuli2014effects,franz2015universal}. In the
jamming limit $p\to 0$ for $\eta>0$, $D(\omega)$ always exhibits the
plateau for small $\omega$, which is fully consistent with previous
numerical simulations of weakly disordered crystals near the jamming
transition point~\cite{goodrich2014solids,charbonneau2019}.

\begin{figure}[t]
\begin{center}
\includegraphics[width=8cm]{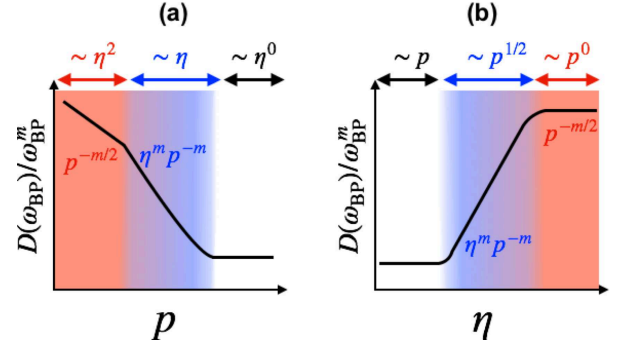} \caption{Scaling of the
boson peak.  (a) and (b) show the $p$ and $\eta$ dependence of the boson
peak intensities, respectively.  } \label{211039_25Oct19}
\end{center}
\end{figure}
Now we want to calculate the boson peak.  For comparison with numerical
simulations, we consider the height of $D(\omega)/\omega^{m}$ at its
peak $\omega=\omega_{\rm BP}$, where $m=1$ and $m=2$ correspond to the
Debye predictions in two and three spatial dimensions,
respectively. Using the scaling of $\delta z$ and
(\ref{122658_26Oct19}), one can deduce the asymptotic behavior for
$m\leq 2$~\footnote{For $m>2$, $D(\omega_{\rm BP})/\omega_{\rm BP}^{m}$
diverges in the RSB phase.} as a function of $p$:
\begin{align}
 \frac{D(\omega_{\rm BP})}{\omega_{\rm BP}^{m}} \sim
 \begin{cases}
  {\rm constant} & \eta\ll p \ll 1\\
  \eta^{m}p^{-m} & \eta^2 \ll p \ll \eta\\
  p^{-\frac{m}{2}} & p \ll \eta^2,
 \end{cases}
 \label{123148_26Oct19}
\end{align}
Eq.~(\ref{123148_26Oct19}) suggests that the boson peak intensity
diverges in the jamming limit $p\to 0$. This scaling is the same of that
of amorphous solids near the jamming transition point observed by a
numerical simulation of three dimensional harmonic
spheres~\cite{mizuno2017}. Repeating the similar calculation, one can
derive the scaling of the boson peak intensity as a function of $\eta$:
\begin{align}
 \frac{D(\omega_{\rm BP})}{\omega_{\rm BP}^{m}} \sim
 \begin{cases}
  p^{-\frac{m}{2}} & \eta\sim 1\\
  \eta^{m}p^{-m} & p \ll \eta \ll p^{1/2}\\
  {\rm constant} & \eta\ll p,
 \end{cases}\label{093646_12Dec19}
\end{align} 
Eq.~(\ref{093646_12Dec19}) suggests that, on the increase of the
polydispersity $\eta$, the boson peak begins to increase at $\eta \sim
p$. This is consistent with a previous numerical simulation of crystals
with small polydispersity~\cite{tong2015crystals}. In
Fig.~\ref{211039_25Oct19}, we summarize the scaling of the boson peak
intensity predicted by the above equations. It is interesting to test
the full scaling behavior by experiments and numerical simulations.

\section{Summary and discussion}
\label{174532_15Jul20}
In this work, we have introduced a mean field model to describe the
jamming transition of crystals with small polydispersity. We solved the
model by using the replica method and determined the full scaling
behaviors of the contact number and density of states above the jamming
transition point. The results are well agreed with previous numerical
simulations.

Another important quantity to characterize the jamming transition is the
gap distribution~\cite{charbonneau2014fractal}. In
Refs.~\cite{charbonneau2019,tsekenis2020jamming}, it is shown that the
gap distribution of the disordered crystal has a different critical
exponent from both perfect crystals and amorphous solids. It is an
interesting future work to see if our model can explain this intriguing
behavior of the gap distribution.

\begin{acknowledgments}
We thank G.~Tsekenis, P.~Urbani, and F.~Zamponi for kind discussions.
We thank P.~Charbonneau for useful comments. This project has received
funding from the European Research Council (ERC) under the European
Union's Horizon 2020 research and innovation program (grant agreement
n.~723955-GlassUniversality) and JSPS KAKENHI Grant Number JP20J00289.
\end{acknowledgments}

\bibliography{apssamp}

\begin{thebibliography}{37}%
\makeatletter
\providecommand \@ifxundefined [1]{%
 \@ifx{#1\undefined}
}%
\providecommand \@ifnum [1]{%
 \ifnum #1\expandafter \@firstoftwo
 \else \expandafter \@secondoftwo
 \fi
}%
\providecommand \@ifx [1]{%
 \ifx #1\expandafter \@firstoftwo
 \else \expandafter \@secondoftwo
 \fi
}%
\providecommand \natexlab [1]{#1}%
\providecommand \enquote  [1]{``#1''}%
\providecommand \bibnamefont  [1]{#1}%
\providecommand \bibfnamefont [1]{#1}%
\providecommand \citenamefont [1]{#1}%
\providecommand \href@noop [0]{\@secondoftwo}%
\providecommand \href [0]{\begingroup \@sanitize@url \@href}%
\providecommand \@href[1]{\@@startlink{#1}\@@href}%
\providecommand \@@href[1]{\endgroup#1\@@endlink}%
\providecommand \@sanitize@url [0]{\catcode `\\12\catcode `\$12\catcode
  `\&12\catcode `\#12\catcode `\^12\catcode `\_12\catcode `\%12\relax}%
\providecommand \@@startlink[1]{}%
\providecommand \@@endlink[0]{}%
\providecommand \url  [0]{\begingroup\@sanitize@url \@url }%
\providecommand \@url [1]{\endgroup\@href {#1}{\urlprefix }}%
\providecommand \urlprefix  [0]{URL }%
\providecommand \Eprint [0]{\href }%
\providecommand \doibase [0]{http://dx.doi.org/}%
\providecommand \selectlanguage [0]{\@gobble}%
\providecommand \bibinfo  [0]{\@secondoftwo}%
\providecommand \bibfield  [0]{\@secondoftwo}%
\providecommand \translation [1]{[#1]}%
\providecommand \BibitemOpen [0]{}%
\providecommand \bibitemStop [0]{}%
\providecommand \bibitemNoStop [0]{.\EOS\space}%
\providecommand \EOS [0]{\spacefactor3000\relax}%
\providecommand \BibitemShut  [1]{\csname bibitem#1\endcsname}%
\let\auto@bib@innerbib\@empty
\bibitem [{\citenamefont {Kittel}\ \emph {et~al.}(1976)\citenamefont {Kittel}
  \emph {et~al.}}]{kittel1976introduction}%
  \BibitemOpen
  \bibfield  {author} {\bibinfo {author} {\bibfnamefont {C.}~\bibnamefont
  {Kittel}} \emph {et~al.},\ }\href@noop {} {\emph {\bibinfo {title}
  {Introduction to solid state physics: 8th Edition}}}\ (\bibinfo  {publisher}
  {Wiley New York},\ \bibinfo {year} {1976})\BibitemShut {NoStop}%
\bibitem [{\citenamefont {Buchenau}\ \emph {et~al.}(1984)\citenamefont
  {Buchenau}, \citenamefont {N\"ucker},\ and\ \citenamefont
  {Dianoux}}]{buchenau1984}%
  \BibitemOpen
  \bibfield  {author} {\bibinfo {author} {\bibfnamefont {U.}~\bibnamefont
  {Buchenau}}, \bibinfo {author} {\bibfnamefont {N.}~\bibnamefont {N\"ucker}},
  \ and\ \bibinfo {author} {\bibfnamefont {A.~J.}\ \bibnamefont {Dianoux}},\
  }\href {\doibase 10.1103/PhysRevLett.53.2316} {\bibfield  {journal} {\bibinfo
   {journal} {Phys. Rev. Lett.}\ }\textbf {\bibinfo {volume} {53}},\ \bibinfo
  {pages} {2316} (\bibinfo {year} {1984})}\BibitemShut {NoStop}%
\bibitem [{\citenamefont {Malinovsky}\ and\ \citenamefont
  {Sokolov}(1986)}]{malinovsky1986}%
  \BibitemOpen
  \bibfield  {author} {\bibinfo {author} {\bibfnamefont {V.}~\bibnamefont
  {Malinovsky}}\ and\ \bibinfo {author} {\bibfnamefont {A.}~\bibnamefont
  {Sokolov}},\ }\href@noop {} {\bibfield  {journal} {\bibinfo  {journal} {Solid
  State Commun.}\ }\textbf {\bibinfo {volume} {57}},\ \bibinfo {pages} {757}
  (\bibinfo {year} {1986})}\BibitemShut {NoStop}%
\bibitem [{\citenamefont {Grigera}\ \emph {et~al.}(2003)\citenamefont
  {Grigera}, \citenamefont {Martin-Mayor}, \citenamefont {Parisi},\ and\
  \citenamefont {Verrocchio}}]{grigera2003phonon}%
  \BibitemOpen
  \bibfield  {author} {\bibinfo {author} {\bibfnamefont {T.}~\bibnamefont
  {Grigera}}, \bibinfo {author} {\bibfnamefont {V.}~\bibnamefont
  {Martin-Mayor}}, \bibinfo {author} {\bibfnamefont {G.}~\bibnamefont
  {Parisi}}, \ and\ \bibinfo {author} {\bibfnamefont {P.}~\bibnamefont
  {Verrocchio}},\ }\href@noop {} {\bibfield  {journal} {\bibinfo  {journal}
  {Nature}\ }\textbf {\bibinfo {volume} {422}},\ \bibinfo {pages} {289}
  (\bibinfo {year} {2003})}\BibitemShut {NoStop}%
\bibitem [{\citenamefont {Shintani}\ and\ \citenamefont
  {Tanaka}(2008)}]{shintani2008universal}%
  \BibitemOpen
  \bibfield  {author} {\bibinfo {author} {\bibfnamefont {H.}~\bibnamefont
  {Shintani}}\ and\ \bibinfo {author} {\bibfnamefont {H.}~\bibnamefont
  {Tanaka}},\ }\href@noop {} {\bibfield  {journal} {\bibinfo  {journal} {Nat.
  Mater.}\ }\textbf {\bibinfo {volume} {7}},\ \bibinfo {pages} {870} (\bibinfo
  {year} {2008})}\BibitemShut {NoStop}%
\bibitem [{\citenamefont {Kaya}\ \emph {et~al.}(2010)\citenamefont {Kaya},
  \citenamefont {Green}, \citenamefont {Maloney},\ and\ \citenamefont
  {Islam}}]{kaya2010}%
  \BibitemOpen
  \bibfield  {author} {\bibinfo {author} {\bibfnamefont {D.}~\bibnamefont
  {Kaya}}, \bibinfo {author} {\bibfnamefont {N.}~\bibnamefont {Green}},
  \bibinfo {author} {\bibfnamefont {C.}~\bibnamefont {Maloney}}, \ and\
  \bibinfo {author} {\bibfnamefont {M.}~\bibnamefont {Islam}},\ }\href@noop {}
  {\bibfield  {journal} {\bibinfo  {journal} {Science}\ }\textbf {\bibinfo
  {volume} {329}},\ \bibinfo {pages} {656} (\bibinfo {year}
  {2010})}\BibitemShut {NoStop}%
\bibitem [{\citenamefont {Phillips}\ and\ \citenamefont
  {Anderson}(1981)}]{phillips1981amorphous}%
  \BibitemOpen
  \bibfield  {author} {\bibinfo {author} {\bibfnamefont {W.~A.}\ \bibnamefont
  {Phillips}}\ and\ \bibinfo {author} {\bibfnamefont {A.}~\bibnamefont
  {Anderson}},\ }\href@noop {} {\emph {\bibinfo {title} {Amorphous solids:
  low-temperature properties}}},\ Vol.~\bibinfo {volume} {24}\ (\bibinfo
  {publisher} {Springer},\ \bibinfo {year} {1981})\BibitemShut {NoStop}%
\bibitem [{\citenamefont {van Hecke}(2009)}]{van2009jamming}%
  \BibitemOpen
  \bibfield  {author} {\bibinfo {author} {\bibfnamefont {M.}~\bibnamefont {van
  Hecke}},\ }\href@noop {} {\bibfield  {journal} {\bibinfo  {journal} {J. Phys.
  Condens. Matter}\ }\textbf {\bibinfo {volume} {22}},\ \bibinfo {pages}
  {033101} (\bibinfo {year} {2009})}\BibitemShut {NoStop}%
\bibitem [{\citenamefont {O'Hern}\ \emph {et~al.}(2003)\citenamefont {O'Hern},
  \citenamefont {Silbert}, \citenamefont {Liu},\ and\ \citenamefont
  {Nagel}}]{ohern2003}%
  \BibitemOpen
  \bibfield  {author} {\bibinfo {author} {\bibfnamefont {C.~S.}\ \bibnamefont
  {O'Hern}}, \bibinfo {author} {\bibfnamefont {L.~E.}\ \bibnamefont {Silbert}},
  \bibinfo {author} {\bibfnamefont {A.~J.}\ \bibnamefont {Liu}}, \ and\
  \bibinfo {author} {\bibfnamefont {S.~R.}\ \bibnamefont {Nagel}},\ }\href
  {\doibase 10.1103/PhysRevE.68.011306} {\bibfield  {journal} {\bibinfo
  {journal} {Phys. Rev. E}\ }\textbf {\bibinfo {volume} {68}},\ \bibinfo
  {pages} {011306} (\bibinfo {year} {2003})}\BibitemShut {NoStop}%
\bibitem [{\citenamefont {Wyart}\ \emph {et~al.}(2005)\citenamefont {Wyart},
  \citenamefont {Silbert}, \citenamefont {Nagel},\ and\ \citenamefont
  {Witten}}]{wyart2005}%
  \BibitemOpen
  \bibfield  {author} {\bibinfo {author} {\bibfnamefont {M.}~\bibnamefont
  {Wyart}}, \bibinfo {author} {\bibfnamefont {L.~E.}\ \bibnamefont {Silbert}},
  \bibinfo {author} {\bibfnamefont {S.~R.}\ \bibnamefont {Nagel}}, \ and\
  \bibinfo {author} {\bibfnamefont {T.~A.}\ \bibnamefont {Witten}},\ }\href
  {\doibase 10.1103/PhysRevE.72.051306} {\bibfield  {journal} {\bibinfo
  {journal} {Phys. Rev. E}\ }\textbf {\bibinfo {volume} {72}},\ \bibinfo
  {pages} {051306} (\bibinfo {year} {2005})}\BibitemShut {NoStop}%
\bibitem [{\citenamefont {Wyart}(2005)}]{wyart2005rigidity}%
  \BibitemOpen
  \bibfield  {author} {\bibinfo {author} {\bibfnamefont {M.}~\bibnamefont
  {Wyart}},\ }\href@noop {} {\bibfield  {journal} {\bibinfo  {journal} {arXiv
  preprint cond-mat/0512155}\ } (\bibinfo {year} {2005})}\BibitemShut {NoStop}%
\bibitem [{\citenamefont {Charbonneau}\ \emph {et~al.}(2014)\citenamefont
  {Charbonneau}, \citenamefont {Kurchan}, \citenamefont {Parisi}, \citenamefont
  {Urbani},\ and\ \citenamefont {Zamponi}}]{charbonneau2014fractal}%
  \BibitemOpen
  \bibfield  {author} {\bibinfo {author} {\bibfnamefont {P.}~\bibnamefont
  {Charbonneau}}, \bibinfo {author} {\bibfnamefont {J.}~\bibnamefont
  {Kurchan}}, \bibinfo {author} {\bibfnamefont {G.}~\bibnamefont {Parisi}},
  \bibinfo {author} {\bibfnamefont {P.}~\bibnamefont {Urbani}}, \ and\ \bibinfo
  {author} {\bibfnamefont {F.}~\bibnamefont {Zamponi}},\ }\href@noop {}
  {\bibfield  {journal} {\bibinfo  {journal} {Nat. Commun.}\ }\textbf {\bibinfo
  {volume} {5}},\ \bibinfo {pages} {3725} (\bibinfo {year} {2014})}\BibitemShut
  {NoStop}%
\bibitem [{\citenamefont {Charbonneau}\ \emph {et~al.}(2017)\citenamefont
  {Charbonneau}, \citenamefont {Kurchan}, \citenamefont {Parisi}, \citenamefont
  {Urbani},\ and\ \citenamefont {Zamponi}}]{charbonneau2017glass}%
  \BibitemOpen
  \bibfield  {author} {\bibinfo {author} {\bibfnamefont {P.}~\bibnamefont
  {Charbonneau}}, \bibinfo {author} {\bibfnamefont {J.}~\bibnamefont
  {Kurchan}}, \bibinfo {author} {\bibfnamefont {G.}~\bibnamefont {Parisi}},
  \bibinfo {author} {\bibfnamefont {P.}~\bibnamefont {Urbani}}, \ and\ \bibinfo
  {author} {\bibfnamefont {F.}~\bibnamefont {Zamponi}},\ }\href@noop {}
  {\bibfield  {journal} {\bibinfo  {journal} {Annu. Rev. Condens. Matter
  Phys.}\ }\textbf {\bibinfo {volume} {8}},\ \bibinfo {pages} {265} (\bibinfo
  {year} {2017})}\BibitemShut {NoStop}%
\bibitem [{\citenamefont {Goodrich}\ \emph {et~al.}(2014)\citenamefont
  {Goodrich}, \citenamefont {Liu},\ and\ \citenamefont
  {Nagel}}]{goodrich2014solids}%
  \BibitemOpen
  \bibfield  {author} {\bibinfo {author} {\bibfnamefont {C.~P.}\ \bibnamefont
  {Goodrich}}, \bibinfo {author} {\bibfnamefont {A.~J.}\ \bibnamefont {Liu}}, \
  and\ \bibinfo {author} {\bibfnamefont {S.~R.}\ \bibnamefont {Nagel}},\
  }\href@noop {} {\bibfield  {journal} {\bibinfo  {journal} {Nat. Phys.}\
  }\textbf {\bibinfo {volume} {10}},\ \bibinfo {pages} {578} (\bibinfo {year}
  {2014})}\BibitemShut {NoStop}%
\bibitem [{\citenamefont {Tong}\ \emph {et~al.}(2015)\citenamefont {Tong},
  \citenamefont {Tan},\ and\ \citenamefont {Xu}}]{tong2015crystals}%
  \BibitemOpen
  \bibfield  {author} {\bibinfo {author} {\bibfnamefont {H.}~\bibnamefont
  {Tong}}, \bibinfo {author} {\bibfnamefont {P.}~\bibnamefont {Tan}}, \ and\
  \bibinfo {author} {\bibfnamefont {N.}~\bibnamefont {Xu}},\ }\href@noop {}
  {\bibfield  {journal} {\bibinfo  {journal} {Sci. Rep.}\ }\textbf {\bibinfo
  {volume} {5}},\ \bibinfo {pages} {15378} (\bibinfo {year}
  {2015})}\BibitemShut {NoStop}%
\bibitem [{\citenamefont {Maxwell}(1864)}]{maxwell1864}%
  \BibitemOpen
  \bibfield  {author} {\bibinfo {author} {\bibfnamefont {J.~C.}\ \bibnamefont
  {Maxwell}},\ }\href@noop {} {\bibfield  {journal} {\bibinfo  {journal} {The
  London, Edinburgh, and Dublin Philosophical Magazine and Journal of Science}\
  }\textbf {\bibinfo {volume} {27}},\ \bibinfo {pages} {250} (\bibinfo {year}
  {1864})}\BibitemShut {NoStop}%
\bibitem [{\citenamefont {Alexander}(1998)}]{alexander1998}%
  \BibitemOpen
  \bibfield  {author} {\bibinfo {author} {\bibfnamefont {S.}~\bibnamefont
  {Alexander}},\ }\href@noop {} {\bibfield  {journal} {\bibinfo  {journal}
  {Phys. Rep.}\ }\textbf {\bibinfo {volume} {296}},\ \bibinfo {pages} {65}
  (\bibinfo {year} {1998})}\BibitemShut {NoStop}%
\bibitem [{\citenamefont {Mizuno}\ \emph {et~al.}(2013)\citenamefont {Mizuno},
  \citenamefont {Mossa},\ and\ \citenamefont {Barrat}}]{mizuno2013elastic}%
  \BibitemOpen
  \bibfield  {author} {\bibinfo {author} {\bibfnamefont {H.}~\bibnamefont
  {Mizuno}}, \bibinfo {author} {\bibfnamefont {S.}~\bibnamefont {Mossa}}, \
  and\ \bibinfo {author} {\bibfnamefont {J.-L.}\ \bibnamefont {Barrat}},\
  }\href@noop {} {\bibfield  {journal} {\bibinfo  {journal} {EPL}\ }\textbf
  {\bibinfo {volume} {104}},\ \bibinfo {pages} {56001} (\bibinfo {year}
  {2013})}\BibitemShut {NoStop}%
\bibitem [{\citenamefont {Guo-Hua}\ \emph {et~al.}(2014)\citenamefont
  {Guo-Hua}, \citenamefont {Qi-Cheng}, \citenamefont {Zhi-Ping}, \citenamefont
  {Xu}, \citenamefont {Qiang},\ and\ \citenamefont {Feng}}]{guo2014effect}%
  \BibitemOpen
  \bibfield  {author} {\bibinfo {author} {\bibfnamefont {Z.}~\bibnamefont
  {Guo-Hua}}, \bibinfo {author} {\bibfnamefont {S.}~\bibnamefont {Qi-Cheng}},
  \bibinfo {author} {\bibfnamefont {S.}~\bibnamefont {Zhi-Ping}}, \bibinfo
  {author} {\bibfnamefont {F.}~\bibnamefont {Xu}}, \bibinfo {author}
  {\bibfnamefont {G.}~\bibnamefont {Qiang}}, \ and\ \bibinfo {author}
  {\bibfnamefont {J.}~\bibnamefont {Feng}},\ }\href@noop {} {\bibfield
  {journal} {\bibinfo  {journal} {Chin. Phys. B}\ }\textbf {\bibinfo {volume}
  {23}},\ \bibinfo {pages} {076301} (\bibinfo {year} {2014})}\BibitemShut
  {NoStop}%
\bibitem [{\citenamefont {Mari}\ \emph {et~al.}(2009)\citenamefont {Mari},
  \citenamefont {Krzakala},\ and\ \citenamefont {Kurchan}}]{mari2009}%
  \BibitemOpen
  \bibfield  {author} {\bibinfo {author} {\bibfnamefont {R.}~\bibnamefont
  {Mari}}, \bibinfo {author} {\bibfnamefont {F.}~\bibnamefont {Krzakala}}, \
  and\ \bibinfo {author} {\bibfnamefont {J.}~\bibnamefont {Kurchan}},\ }\href
  {\doibase 10.1103/PhysRevLett.103.025701} {\bibfield  {journal} {\bibinfo
  {journal} {Phys. Rev. Lett.}\ }\textbf {\bibinfo {volume} {103}},\ \bibinfo
  {pages} {025701} (\bibinfo {year} {2009})}\BibitemShut {NoStop}%
\bibitem [{\citenamefont {Georges}\ \emph {et~al.}(1996)\citenamefont
  {Georges}, \citenamefont {Kotliar}, \citenamefont {Krauth},\ and\
  \citenamefont {Rozenberg}}]{georges1996}%
  \BibitemOpen
  \bibfield  {author} {\bibinfo {author} {\bibfnamefont {A.}~\bibnamefont
  {Georges}}, \bibinfo {author} {\bibfnamefont {G.}~\bibnamefont {Kotliar}},
  \bibinfo {author} {\bibfnamefont {W.}~\bibnamefont {Krauth}}, \ and\ \bibinfo
  {author} {\bibfnamefont {M.~J.}\ \bibnamefont {Rozenberg}},\ }\href {\doibase
  10.1103/RevModPhys.68.13} {\bibfield  {journal} {\bibinfo  {journal} {Rev.
  Mod. Phys.}\ }\textbf {\bibinfo {volume} {68}},\ \bibinfo {pages} {13}
  (\bibinfo {year} {1996})}\BibitemShut {NoStop}%
\bibitem [{\citenamefont {Parisi}\ and\ \citenamefont
  {Zamponi}(2010)}]{parisi2010mean}%
  \BibitemOpen
  \bibfield  {author} {\bibinfo {author} {\bibfnamefont {G.}~\bibnamefont
  {Parisi}}\ and\ \bibinfo {author} {\bibfnamefont {F.}~\bibnamefont
  {Zamponi}},\ }\href@noop {} {\bibfield  {journal} {\bibinfo  {journal} {Rev.
  Mod. Phys.}\ }\textbf {\bibinfo {volume} {82}},\ \bibinfo {pages} {789}
  (\bibinfo {year} {2010})}\BibitemShut {NoStop}%
\bibitem [{\citenamefont {Frisch}\ and\ \citenamefont
  {Percus}(1999)}]{frisch1999}%
  \BibitemOpen
  \bibfield  {author} {\bibinfo {author} {\bibfnamefont {H.~L.}\ \bibnamefont
  {Frisch}}\ and\ \bibinfo {author} {\bibfnamefont {J.~K.}\ \bibnamefont
  {Percus}},\ }\href {\doibase 10.1103/PhysRevE.60.2942} {\bibfield  {journal}
  {\bibinfo  {journal} {Phys. Rev. E}\ }\textbf {\bibinfo {volume} {60}},\
  \bibinfo {pages} {2942} (\bibinfo {year} {1999})}\BibitemShut {NoStop}%
\bibitem [{\citenamefont {M{\'e}zard}\ \emph {et~al.}(1987)\citenamefont
  {M{\'e}zard}, \citenamefont {Parisi},\ and\ \citenamefont
  {Virasoro}}]{mezard1987spin}%
  \BibitemOpen
  \bibfield  {author} {\bibinfo {author} {\bibfnamefont {M.}~\bibnamefont
  {M{\'e}zard}}, \bibinfo {author} {\bibfnamefont {G.}~\bibnamefont {Parisi}},
  \ and\ \bibinfo {author} {\bibfnamefont {M.}~\bibnamefont {Virasoro}},\
  }\href@noop {} {\emph {\bibinfo {title} {Spin glass theory and beyond: An
  Introduction to the Replica Method and Its Applications}}},\ Vol.~\bibinfo
  {volume} {9}\ (\bibinfo  {publisher} {World Scientific Publishing Company},\
  \bibinfo {year} {1987})\BibitemShut {NoStop}%
\bibitem [{Note1()}]{Note1}%
  \BibitemOpen
  \bibinfo {note} {This is a slightly different condition from particle systems
  in $d$ spatial dimensions where all particles are mobile, and the isostatic
  condition leads to $Z_{\protect \rm iso}=2d$~\cite
  {van2009jamming}.}\BibitemShut {Stop}%
\bibitem [{\citenamefont {Franz}\ \emph {et~al.}(2015)\citenamefont {Franz},
  \citenamefont {Parisi}, \citenamefont {Urbani},\ and\ \citenamefont
  {Zamponi}}]{franz2015universal}%
  \BibitemOpen
  \bibfield  {author} {\bibinfo {author} {\bibfnamefont {S.}~\bibnamefont
  {Franz}}, \bibinfo {author} {\bibfnamefont {G.}~\bibnamefont {Parisi}},
  \bibinfo {author} {\bibfnamefont {P.}~\bibnamefont {Urbani}}, \ and\ \bibinfo
  {author} {\bibfnamefont {F.}~\bibnamefont {Zamponi}},\ }\href@noop {}
  {\bibfield  {journal} {\bibinfo  {journal} {PNAS}\ }\textbf {\bibinfo
  {volume} {112}},\ \bibinfo {pages} {14539} (\bibinfo {year}
  {2015})}\BibitemShut {NoStop}%
\bibitem [{\citenamefont {Livan}\ \emph {et~al.}(2018)\citenamefont {Livan},
  \citenamefont {Novaes},\ and\ \citenamefont {Vivo}}]{livan2018introduction}%
  \BibitemOpen
  \bibfield  {author} {\bibinfo {author} {\bibfnamefont {G.}~\bibnamefont
  {Livan}}, \bibinfo {author} {\bibfnamefont {M.}~\bibnamefont {Novaes}}, \
  and\ \bibinfo {author} {\bibfnamefont {P.}~\bibnamefont {Vivo}},\ }\href@noop
  {} {\emph {\bibinfo {title} {Introduction to random matrices: theory and
  practice}}}\ (\bibinfo  {publisher} {Springer},\ \bibinfo {year}
  {2018})\BibitemShut {NoStop}%
\bibitem [{\citenamefont {Franz}\ \emph {et~al.}(2017)\citenamefont {Franz},
  \citenamefont {Parisi}, \citenamefont {Sevelev}, \citenamefont {Urbani},
  \citenamefont {Zamponi},\ and\ \citenamefont
  {Sevelev}}]{franz2017universality}%
  \BibitemOpen
  \bibfield  {author} {\bibinfo {author} {\bibfnamefont {S.}~\bibnamefont
  {Franz}}, \bibinfo {author} {\bibfnamefont {G.}~\bibnamefont {Parisi}},
  \bibinfo {author} {\bibfnamefont {M.}~\bibnamefont {Sevelev}}, \bibinfo
  {author} {\bibfnamefont {P.}~\bibnamefont {Urbani}}, \bibinfo {author}
  {\bibfnamefont {F.}~\bibnamefont {Zamponi}}, \ and\ \bibinfo {author}
  {\bibfnamefont {M.}~\bibnamefont {Sevelev}},\ }\href@noop {} {\bibfield
  {journal} {\bibinfo  {journal} {SciPost Physics}\ }\textbf {\bibinfo {volume}
  {2}},\ \bibinfo {pages} {019} (\bibinfo {year} {2017})}\BibitemShut {NoStop}%
\bibitem [{Note2()}]{Note2}%
  \BibitemOpen
  \bibinfo {note} {The condition of the isostaticity is now $z=z_{\protect \rm
  iso}\equiv Z_{\protect \rm iso}/d=1$.}\BibitemShut {Stop}%
\bibitem [{\citenamefont {Franz}\ and\ \citenamefont
  {Parisi}(2016)}]{franz2016simplest}%
  \BibitemOpen
  \bibfield  {author} {\bibinfo {author} {\bibfnamefont {S.}~\bibnamefont
  {Franz}}\ and\ \bibinfo {author} {\bibfnamefont {G.}~\bibnamefont {Parisi}},\
  }\href@noop {} {\bibfield  {journal} {\bibinfo  {journal} {J. Phys. A}\
  }\textbf {\bibinfo {volume} {49}},\ \bibinfo {pages} {145001} (\bibinfo
  {year} {2016})}\BibitemShut {NoStop}%
\bibitem [{\citenamefont {Acharya}\ \emph {et~al.}(2020)\citenamefont
  {Acharya}, \citenamefont {Sengupta}, \citenamefont {Chakraborty},\ and\
  \citenamefont {Ramola}}]{acharya2019athermal}%
  \BibitemOpen
  \bibfield  {author} {\bibinfo {author} {\bibfnamefont {P.}~\bibnamefont
  {Acharya}}, \bibinfo {author} {\bibfnamefont {S.}~\bibnamefont {Sengupta}},
  \bibinfo {author} {\bibfnamefont {B.}~\bibnamefont {Chakraborty}}, \ and\
  \bibinfo {author} {\bibfnamefont {K.}~\bibnamefont {Ramola}},\ }\href
  {\doibase 10.1103/PhysRevLett.124.168004} {\bibfield  {journal} {\bibinfo
  {journal} {Phys. Rev. Lett.}\ }\textbf {\bibinfo {volume} {124}},\ \bibinfo
  {pages} {168004} (\bibinfo {year} {2020})}\BibitemShut {NoStop}%
\bibitem [{Note3()}]{Note3}%
  \BibitemOpen
  \bibinfo {note} {In finite $d$, $D(\omega )$ for $\omega <\omega _0$ is
  described by the Debye theory $D(\omega )\sim \omega ^{d-1}$.}\BibitemShut
  {Stop}%
\bibitem [{\citenamefont {DeGiuli}\ \emph {et~al.}(2014)\citenamefont
  {DeGiuli}, \citenamefont {Laversanne-Finot}, \citenamefont {D{\"u}ring},
  \citenamefont {Lerner},\ and\ \citenamefont {Wyart}}]{degiuli2014effects}%
  \BibitemOpen
  \bibfield  {author} {\bibinfo {author} {\bibfnamefont {E.}~\bibnamefont
  {DeGiuli}}, \bibinfo {author} {\bibfnamefont {A.}~\bibnamefont
  {Laversanne-Finot}}, \bibinfo {author} {\bibfnamefont {G.}~\bibnamefont
  {D{\"u}ring}}, \bibinfo {author} {\bibfnamefont {E.}~\bibnamefont {Lerner}},
  \ and\ \bibinfo {author} {\bibfnamefont {M.}~\bibnamefont {Wyart}},\
  }\href@noop {} {\bibfield  {journal} {\bibinfo  {journal} {Soft Matter}\
  }\textbf {\bibinfo {volume} {10}},\ \bibinfo {pages} {5628} (\bibinfo {year}
  {2014})}\BibitemShut {NoStop}%
\bibitem [{\citenamefont {Charbonneau}\ \emph {et~al.}(2019)\citenamefont
  {Charbonneau}, \citenamefont {Corwin}, \citenamefont {Fu}, \citenamefont
  {Tsekenis},\ and\ \citenamefont {van~der Naald}}]{charbonneau2019}%
  \BibitemOpen
  \bibfield  {author} {\bibinfo {author} {\bibfnamefont {P.}~\bibnamefont
  {Charbonneau}}, \bibinfo {author} {\bibfnamefont {E.~I.}\ \bibnamefont
  {Corwin}}, \bibinfo {author} {\bibfnamefont {L.}~\bibnamefont {Fu}}, \bibinfo
  {author} {\bibfnamefont {G.}~\bibnamefont {Tsekenis}}, \ and\ \bibinfo
  {author} {\bibfnamefont {M.}~\bibnamefont {van~der Naald}},\ }\href {\doibase
  10.1103/PhysRevE.99.020901} {\bibfield  {journal} {\bibinfo  {journal} {Phys.
  Rev. E}\ }\textbf {\bibinfo {volume} {99}},\ \bibinfo {pages} {020901}
  (\bibinfo {year} {2019})}\BibitemShut {NoStop}%
\bibitem [{Note4()}]{Note4}%
  \BibitemOpen
  \bibinfo {note} {For $m>2$, $D(\omega _{\protect \rm BP})/\omega _{\protect
  \rm BP}^{m}$ diverges in the RSB phase.}\BibitemShut {Stop}%
\bibitem [{\citenamefont {Mizuno}\ \emph {et~al.}(2017)\citenamefont {Mizuno},
  \citenamefont {Shiba},\ and\ \citenamefont {Ikeda}}]{mizuno2017}%
  \BibitemOpen
  \bibfield  {author} {\bibinfo {author} {\bibfnamefont {H.}~\bibnamefont
  {Mizuno}}, \bibinfo {author} {\bibfnamefont {H.}~\bibnamefont {Shiba}}, \
  and\ \bibinfo {author} {\bibfnamefont {A.}~\bibnamefont {Ikeda}},\
  }\href@noop {} {\bibfield  {journal} {\bibinfo  {journal} {PNAS}\ }\textbf
  {\bibinfo {volume} {114}},\ \bibinfo {pages} {E9767} (\bibinfo {year}
  {2017})}\BibitemShut {NoStop}%
\bibitem [{\citenamefont {Tsekenis}(2020)}]{tsekenis2020jamming}%
  \BibitemOpen
  \bibfield  {author} {\bibinfo {author} {\bibfnamefont {G.}~\bibnamefont
  {Tsekenis}},\ }\href@noop {} {\bibfield  {journal} {\bibinfo  {journal}
  {arXiv preprint arXiv:2006.07373}\ } (\bibinfo {year} {2020})}\BibitemShut
  {NoStop}%
\end{thebibliography}%

\onecolumngrid
\appendix

\section{Interaction potential in the large dimensional limit}
\label{203409_15Jul20}

Our model consists of a tracer particle and $M$ nearest neighbor (NN)
particles.  The tracer is located \textit{in} a $d$-dimensional
hypersphere, and the $M$ NN particles are fixed on the surface of the
hypersphere.  The interaction potential is given by
\begin{align}
 V(\bX) = \sum_{\mu=1}^M v(h^\mu),\label{093640_9Apr20}
 \end{align}
where $v(h)=h^2\theta(-h)/2$ and 
\begin{align}
 h^\mu = \sqrt{d}\left(\abs{\bX-\by^\mu}-\sigma^\mu\right).\label{103553_8Apr20}
\end{align}
Here the pre-factor $\sqrt{d}$ is necessary to keep $V(\bX)=O(d^0)$ in the
$d\to\infty$ limit, as we will see below. $\bX=\{X_1,\cdots, X_d\}$ and
$\by^\mu=\{y_1^\mu,\cdots,y_d^\mu\}$ denote the positions of the tracer
and $\mu$-th NN, respectively. The distribution of $\by^\mu$ is 
\begin{align}
 P(\by^\mu) = 
 \frac{\delta(\by^\mu\cdot\by^\mu-d)}{\int d\by^\mu\delta(\by^\mu\cdot\by^\mu-d)}.
\end{align} 
$\sigma^\mu$ denotes the interaction range between the tracer and
$\mu$-th obstacles. We consider that $\sigma^\mu$ has the following form
\begin{align}
\sigma^\mu = \sigma\left(1 + \frac{1}{d}b^\mu\right),\label{103342_8Apr20}
\end{align}
where $\sigma$ controls the mean size of the particles, and $b^\mu$ denotes the
polydispersity. $b^\mu$ follows the normal distribution of zero mean and
variance $\eta^2$:
\begin{align}
 P(b^\mu) = \frac{1}{\sqrt{2\eta^2\pi}}\exp\left[-\frac{(b^\mu)^2}{2\eta^2}\right].
\end{align}
For $\eta=0$, the jamming transition occurs at $\sigma =
\sigma_J^0\equiv \sqrt{d}$ at which $\bX=0$ and $\gap^\mu=0$ for all
$\mu$. We expand $\sigma$ around $\sigma_J^0$ as 
\begin{align}
 \sigma = \sigma_J^0\left(1+ \frac{1}{d}{a}\right),\label{154840_16May20}
\end{align}
where the pre-factor of $a$, $1/d$, is necessary to keep the relative
interaction volume $(\sigma/\sigma_J^0)^d$ finite in the limit of $d\to\infty$.
Substituting Eqs.~(\ref{103342_8Apr20}) and (\ref{154840_16May20}) into
the gap function $\gap^\mu$ and expanding by $1/d$, we get
\begin{align}
 h^\mu = \sqrt{d}\left[\sqrt{d}\left(1+\frac{\bX\cdot\bX}{2d}-\frac{\bX\cdot\by^\mu}{d}+\cdots\right) -\sigma^\mu\right]
 = d\left(\frac{\bX\cdot\bX}{2d}-\frac{\bX\cdot\by^\mu}{d}
 -\frac{{a}+b^\mu}{d}
 +
 O\left(\frac{\bX\cdot\bX}{d^2},\frac{\bX\cdot\by^\mu}{d^2}\right)\right),
\end{align}
where we have used $\by^\mu\cdot\by^\mu = d$.  We require that the
first-order terms have the same magnitude.  This is possible if the
following conditions are satisfied
\begin{align}
& \bX\cdot\bX = O(1),\label{105954_8Apr20}\\
 & \bX\cdot\by^\mu = O(1).\label{110130_8Apr20}
\end{align}
Eq.~(\ref{105954_8Apr20}) implies that $\sum_{i=1}^d X_i^2 =O(1)$ or
$X_i^2 = O(d^{-1})$.  We introduce a new variable of order one:
\begin{align}
 x_i = \sqrt{d}X_i.\label{083306_9Apr20}
\end{align}
Eqs.~(\ref{110130_8Apr20}) and (\ref{083306_9Apr20}) lead to $\sum_{i=1}^d x_i y_i^\mu =
O(\sqrt{d})$, which is a natural result because $y_i^\mu$ is a
random variable of zero mean and unit variance.
Up to the first order, we get the following result:
\begin{align}
 h^\mu = \frac{\bx\cdot\bx}{2d}
 -\frac{\bx\cdot\by^\mu}{\sqrt{d}}-{a}-b^\mu + O(d^{-1}).
\end{align}

In summary, in the limit of $d\to\infty$, the interaction potential is 
\begin{align}
 V(\bX) &= \sum_{\mu=1}^M v(h^\mu),\label{112241_10Mar20}\new
 h^\mu &= \frac{\bx\cdot\bx}{2d}-\frac{\bx\cdot\by^\mu}{\sqrt{d}}-{a}-b^\mu. 
\end{align}
The gap function $h^\mu$ has a similar form to the perceptron, except
the additional terms $\bx\cdot\bx/2d$ and
$b^\mu$~\cite{franz2016simplest,franz2017universality}. Therefore, we
can apply the same technique of that of the perceptron to investigate
the model.

\section{Free energy}
\label{173040_17Jul20}
Although we only investigate the model at zero temperature $T=0$, we
fist consider the free-energy at finite $T$ to apply the technique of
the statistical mechanics and then take the limit of $T\to 0$. The
free-energy can be written as
\begin{align}
 -\beta F = \lim_{d\to\infty}\frac{1}{d}\left[\log \mathcal{Z}\right]_{\by,b},\label{112334_9Apr20}
\end{align}
where 
\begin{align}
 \left[\bullet \right]_{\by,b} = \prod_{\mu=1}^M \int P(\by^\mu)d\by^\mu \int P(b^\mu)db^\mu\bullet,
\end{align}
and
\begin{align}
 \mathcal{Z} = \int d\bx e^{-\beta V(\bx)} = \int d\bx \prod_{\mu=1}^M
 \left[\int dr^\mu e^{-\beta v(r^\mu+\bx\cdot\bx/2d-{a})}\delta\left(r^\mu-d^{-1/2}\bx\cdot\by^\mu+b^\mu\right)\right].
\end{align}
Here we introduced the inverse temperature $\beta=1/T$.

First, we will show that the average for $y_i^\mu$ can be replaced by
the average for a normal distribution of zero mean and unit variance.
The mean value of $f(\by^\mu) = \log \mathcal{Z}$ is represented as
\begin{align}
\left[f(\by^\mu)\right]_{\by^\mu} \sim \int d\by^\mu \delta(\by^\mu\cdot\by^\mu-d)f(\by^\mu)
 \sim \int d\by^\mu \int d\lambda  e^{-\frac{\lambda}{2}(\by^\mu\cdot\by^\mu-d)}f(\by^\mu)
 = \int d\lambda \exp\left[\frac{\lambda}{2}d -\log\int d\by^\mu
 e^{-\frac{\lambda}{2}\by^\mu\cdot\by^\mu}f(\by^\mu)\right].\label{112126_9Apr20}
\end{align}
In the limit $d\to\infty$, we can evaluate the integral of $\lambda$ by
using the saddle point method. The saddle point condition is
\begin{align}
 d = \frac{\int d\by^\mu e^{-\frac{\lambda}{2}\by^\mu\cdot\by^\mu+\log f(\by^\mu)}\by^\mu\cdot\by^\mu}
 {\int d\by^\mu e^{-\frac{\lambda}{2}\by^\mu\cdot\by^\mu+\log f(\by^\mu)}} = \lambda d 
\to \lambda = 1.
 \label{112140_9Apr20}
\end{align}
where we used $\by^\mu\cdot\by^\mu = O(d)$ and $\log f = \log\log \mathcal{Z} =
O(\log d)\ll d$. Applying the saddle point method for the integral of
$\lambda$ in Eq.~(\ref{112126_9Apr20}), we get
\begin{align}
\left[f(\by^\mu)\right]_{\by^\mu} \sim \int d\by^\mu e^{-\frac{1}{2}\by^\mu\cdot\by^\mu}f(\by^\mu),
\end{align}
meaning that the distribution function of $y_i^\mu$ converges to a
normal distribution of mean zero and unit variance in the limit of
$d\to\infty$. This can greatly simplify the
calculation as we will see below.

Now, we calculate the free energy Eq.~(\ref{112334_9Apr20}) by using the
replica method:
\begin{align}
 -\beta F = \lim_{n\to 0}\lim_{d\to\infty}\frac{\log\left[\mathcal{Z}^n\right]_{\by,b}}{nd}.\label{083921_15Apr20}
\end{align}
Using the Fourier transformation of the delta function $\delta(x) =
(2\pi)^{-1}\int d\hr e^{i\hr x}$, the partition functions can be written
as
\begin{align}
 \left[\mathcal{Z}^n\right]_{\by,b} & \sim
 \int \left(\prod_{a} d\bx^a\right)\left(\prod_{a,\mu}dr_a^\mu d\hr_a^\mu\right)
 \left[e^{\sum_{a\mu}i\hr_a^\mu(r_a^\mu-d^{-1/2}\bx^a\cdot\by^\mu+b^\mu)}\right]_{\by,b}
 \prod_{a\mu}e^{-\beta v(r_a^\mu+\bx\cdot\bx/2d-{a})},\label{123437_14Apr20}
\end{align}
where $a=1,\cdots, n$ and $\mu=1,\cdots, M$.
Since $\by^\mu$ and $b^\mu$ follow the normal distribution, 
one can show that
\begin{align}
\left[e^{-\sum_{a\mu}i\hr_a^\mu (d^{-1/2}\bx^a\cdot\by^\mu - b^\mu)}\right]_{\by,b}
 = e^{-\frac{1}{2}\sum_{ab\mu}\hr_a^\mu\hr_b^\mu (Q_{ab}+\eta^2)},
\end{align}
where we have introduced a new variable
\begin{align}
 Q_{ab} = \frac{1}{d}\bx^a\cdot\bx^b.
\end{align}
The Jacobian of the change of the variables is 
\begin{align}
\prod_{a=1}^n\int d\bx^a = \prod_{ab}\int dQ_{ab}\delta(dQ_{ab}-\bx^a\cdot\bx^b)
 \sim\prod_{ab}\int dQ_{ab}e^{\frac{d}{2}\log \det Q}.
\end{align}
Using those results, Eq.~(\ref{123437_14Apr20}) can be rewritten as 
\begin{align}
 \left[\mathcal{Z}^n\right]_{\by,b} &\sim
 \prod_{ab}\int dQ_{ab} e^{\frac{d}{2}\log\det Q}
 \left[\int \left(\prod_a dr_a d\hr_a\right)e^{\sum_a i\hr_a r_a -\frac{1}{2}\sum_{ab}\hr_a \hr_b (Q_{ab}+\eta^2)
 -\beta v(r_a+Q_{aa}/2-{a})} \right]^M\new
& \sim  \prod_{ab}\int dQ_{ab} e^{\frac{d}{2}\log\det Q}
 \left[\int \left(\prod_a dr_a d\hr_a\right)
 \left.e^{\sum_a i\hr_a r_a
 + \frac{1}{2}\sum_{ab}(Q_{ab}+\eta^2)\pdiff{^2}{k_a\partial k_b}
 - \sum_a i\hr_a k_a
 -\beta v(r_a+Q_{aa}/2-{a})}\right|_{k_a=0}
 \right]^M\new
 &\sim
  \prod_{ab}\int dQ_{ab} e^{\frac{d}{2}\log\det Q}
 \left[
 \left.e^{
 \frac{1}{2}\sum_{ab}(Q_{ab}+\eta^2)\pdiff{^2}{k_a\partial k_b}
 -\beta v(k_a+Q_{aa}/2-{a})}\right|_{k_a=0}
 \right]^M\new
 &\sim e^{d S(Q_{ab}^*)},
\end{align}
where
\begin{align}
 &S(Q_{ab}) = \frac{1}{2}\log\det Q
 + \alpha \log\left[e^{\frac{1}{2}\sum_{ab}(Q_{ab}+\eta^2)\pdiff{^2}{h_a\partial h_b}}
 \prod_{a=1}^n e^{-\beta v(h^a)}\right]_{h^a=Q_{aa}/2-{a}},\new
 &\alpha = \frac{M}{d},
\end{align}
and $Q_{ab}^*$ detenos the saddle-point value satisfying
$\partial_{Q_{ab}^*}S(Q_{ab}^*)=0$. Finally, the free-energy Eq.~(\ref{083921_15Apr20}) is
calculated as
\begin{align}
 -\beta F = \lim_{n\to 0}\frac{S(Q_{ab}^*)}{n}.
\end{align}

\section{Calculation with the replica symmetric Ansatz}
\label{173250_17Jul20}

\subsection{Free energy}
Here we investigate the model by assuming the replica symmetric (RS)
Ansatz:
\begin{align}
 Q_{ab}^{\rm RS} &= \delta_{ab}q_0 + (1-\delta_{ab})q.
\end{align}
Then, the free-energy is 
\begin{align}
 & -\beta F_{\rm RS} = \lim_{n\to 0}\frac{S(Q_{ab}^{\rm RS})}{n}
 = \frac{1}{2}\left[\log(q_0-q) + \frac{q}{q_0-q} \right]
 + \left.\alpha \gamma_{q+\eta^2}*f(q,h)\right|_{h=q_0/2-{a}},\new
 &f(q,h) = \log \gamma_{q_0-q}*e^{-\beta v(h)},\label{084807_15Apr20}
\end{align}
where we used the abbreviations: 
\begin{align}
 & \gamma_q*\bullet(h) = \int_{-\infty}^\infty dh'\gamma_q(h-h')\bullet(h'),\new
 & \gamma_q(h)  = \frac{1}{\sqrt{2\pi q}}e^{-\frac{h^2}{2q}}.
\end{align}
For low $T$, we can perform the harmonic expansion:
\begin{align}
 q_0-q= \frac{1}{2}\ave{(x^a-x^b)^2}  = T\chi + O(T^2).\label{180115_12Mar20}
\end{align}
Substituting it
into the free-energy, Eq.~(\ref{084807_15Apr20}), we get in the $T\to 0$ limit
\begin{align}
&f(q,h) \sim -\frac{\beta h^2}{2(1+\chi)}\theta(-h),\new
& e_{\rm RS} \equiv \lim_{T\to 0}F_{\rm RS} = -\frac{q_0}{2\chi} + \frac{\alpha}{2(1+\chi)}
 \int_{-\infty}^0 dh \gamma_{q_0 + \eta^2}(q_0/2-{a}-h)h^2.
\end{align}
$\chi$ and $q_0$ are determined by the saddle point equations:
\begin{align}
 &\left(1+\frac{1}{\chi}\right)^2 q_0  = \alpha \int_{-\infty}^0 dh 
\gamma_{q_0 + \eta^2}(q_0/2-{a}-h)h^2,\new
 &1+\frac{1}{\chi} =\alpha \pdiff{}{q_0}\int_{-\infty}^0 dh \gamma_{q_0 + \eta^2}(q_0/2-{a}-h)h^2
 =  \alpha \int_{-\infty}^0 dh \gamma_{q_0 + \eta^2}(q_0/2-{a}-h)(1+h).\label{180205_12Mar20}
\end{align}
The equations can be solved numerically, which allows us to calculate
$q_0$ and $\chi$ for given $a$ and $\eta^2$.

\subsection{Gap distribution function}
Our goal is to calculate the contact number $z$ as a function of the
pressure $p$.  For this purpose, it is convenient to introduce the gap
distribution function:
\begin{align}
 g(h) \equiv \frac{1}{d}\ave{\sum_{\mu=1}^M \delta(h-\hh^\mu)},
\end{align}
where $\langle\bullet\rangle$ denotes the average for both thermal
fluctuation and quenched disorder. $g(h)$ can be calculated as
\begin{align}
 g(h) = \fdiff{F_{\rm RS}}{v(h)}
 = \left.\alpha e^{-\beta v(h)}\gamma_{q_0+\eta^2}*e^{-f(q,h)}\gamma_{q_0-q}(h'-h)
  \right|_{h'=q_0/2-{a}}.
\end{align}
In the limit of $T\to 0$, the saddle point method leads to
\begin{align}
 g(h) = 
\begin{cases}
 \alpha (1+\chi)\gamma_{q_0+\eta^2}(q_0/2-{a}-(1+\chi)h) & h\leq 0,\\
 \alpha \gamma_{q_0+\eta^2}(q_0/2-{a}-h) & h> 0.
\end{cases}\label{171602_15Mar20} 
\end{align}
The contact number per degree of freedom $z$ and pressure $p$ are calculated from
$g(h)$ as
\begin{align}
 & z \equiv \frac{1}{d}\sum_{\mu=1}^M\ave{\theta(-\hh^\mu)} = \int_{-\infty}^\infty dh g(h)\theta(-h),\label{171654_15Mar20}\\
& p \equiv -\frac{1}{d}\sum_{\mu=1}^M\ave{v'(\hh^\mu)}= -\int_{-\infty}^\infty dh g(h)\theta(-h)h.\label{171659_15Mar20}
\end{align}

\subsection{Numerics}

We calculate $z$ as a function of $p$ with the following steps:
\begin{enumerate}
 \item Calculate $q_0$ and $\chi$ as functions of $a$ and
       $\eta$ by solving Eqs.~(\ref{180205_12Mar20}).

 \item Calculate $z$ and $p$ by substituting the above results into Eq.~(\ref{171654_15Mar20}) and (\ref{171659_15Mar20}).
 \item Plot $z$ as a function of $p$.
\end{enumerate}
We found that the above algorithm does not converge for $M\sim d$.  This
may imply that if the number of the NN particles is not large enough,
the tracer particle can escape, and the harmonic expansion,
Eq.~(\ref{180115_12Mar20}), breaks down. To avoid this problem, in the
main text, we show the results for $M=10d\gg d$. We checked that the
qualitatively same results are obtained for different values of $M$ as long as $M \gg d$.

\subsection{Scaling}
We derive the scaling behavior near the jamming transition point for
$\eta\ll 1$ from the asymptotics of the RS equations.

First, we discuss the scaling at the jamming transition point.  
At the transition point, the harmonic expansion breaks down, meaning
that $\chi\to\infty$. Therefore, 
Eqs.~(\ref{180205_12Mar20}) reduce to
\begin{align}
  &q_0  = \alpha \int_{-\infty}^0 dh 
\gamma_{q_0 + \eta^2}(q_0/2-{a}-h)h^2,\new
 &1  =  \alpha \int_{-\infty}^0 dh \gamma_{q_0 + \eta^2}(q_0/2-{a}-h)(1+h).\label{052349_17May20}
\end{align}
By solving the above equations, one can calculate $a$ and $q_0$ at
jamming for given $\eta$.
From an asymptotic analysis for $\eta\ll 1$, we can show  that
\begin{align}
a\sim -\eta,\ q_0 \sim \eta^2.\label{064007_17May20}
\end{align}
The results is consistent with a naive dimensional analysis: both $a$
and $\eta$ have the dimension of length, leading to $a\propto \eta$, and
$q_0$ has the dimension of the squared of length, leading to $q_0\propto
a^2 \propto \eta^2$.

To get the scaling above jamming, we rewrite the saddle point equations
Eqs.~(\ref{180205_12Mar20}) as
\begin{align}
& \frac{q_0}{\chi^2} = \int_{-\infty}^0 dh g(h)h^2 = 2e_{\rm RS},\label{063913_17May20}\\
 & z-1 = \frac{1}{\chi} + (1+\chi)p,\label{064048_17May20}
\end{align}
Using Eq.~(\ref{063913_17May20}), we get
\begin{align}
 \chi \sim \sqrt{\frac{q_0}{e_{\rm RS}}} \sim \frac{\eta}{p},
\end{align}
where we used Eq.~(\ref{064007_17May20}) and $e_{\rm RS}\sim
\ave{h^2}\sim \ave{h}^2\sim p^2$.  Substituting it into
Eq.~(\ref{064048_17May20}), we get 
\begin{align}
 z-1 \sim c_1\frac{p}{\eta} + c_2\eta,
\end{align}
where $c_1$ and $c_2$ denote constants.
On the contrary, far from the jamming transition point, we get $z\sim
\alpha$ as the tracer contact with most NN particles. Summarizing 
the results, the RS Ansatz predicts the following scaling
\begin{align}
 z-1 \sim
 \begin{cases}
  \alpha-1 & \eta\ll p,\\
  \eta^{-1}p & \eta^2 \ll p \ll \eta,\\
  \eta & p\ll \eta^2.
 \end{cases}\label{220929_17May20}
\end{align}
Note however that the above result for $p\ll \eta^2$ is incorrect
because the RS solution becomes unstable. As discussed below, the
correct scaling $\delta z\sim p^{1/2}$ is obtained by using the RSB
equations. As dicussed in the main text, if one plots $z-1$ as a
function of $\eta^{-1}p$, the results for $p\gg\eta^2$ collpase on a
single master curve.

\section{Eigenvalue distribution}

We consider the Hessian Matrix:
\begin{align}
 M_{ij} = \pdiff{^2V}{x_i\partial x_j} = \sum_{\mu=1}^M
\left[
 v''(\hh^\mu)\pdiff{h^\mu}{x_i}\pdiff{\hh^\mu}{x_j}
 + v'(\hh^\mu)\pdiff{^2 \hh^\mu}{x_i\partial x_j}
 \right] \approx \frac{1}{d}\sum_{\mu=1}^M \left[y_i^\mu y_j^\mu + \hh^\mu \delta_{ij}\right]\theta(-h^\mu),
\end{align}
where we have dropped the sub-leading terms in the large dimensional
limit. From the central limit theorem, the diagonal term converges to
the pressure:
\begin{align}
 \frac{1}{d}\sum_{\mu=1}^M \theta(-\hh^\mu)\hh^\mu \delta_{ij} \sim -p \delta_{ij}.
\end{align}
In the previous sections, we have shown that $y_i^\mu$ follows the
normal distribution of zero mean and unit variance. Therefore, $M_{ij}$
is a Wishart matrix shifted by $p$~\cite{livan2018introduction}. The
eigenvalue distribution follows the Marchenko–Pastur (MP)
law~\cite{franz2015universal}:
\begin{align}
& \rho(\lambda)  = \frac{1}{2\pi}\frac{\sqrt{(\lambda-\lambda_-)(\lambda_+-\lambda)}}{\lambda+p},\new
& \lambda_{\pm} = \left(\sqrt{z}\pm 1\right)^2 -p.
\end{align}
In particular, we are interested in the minimal eigenvalue $\lambda_- =
(\sqrt{z}-1)^2-p$, which vanishes when the RS Ansatz becomes unstable, and the RSB phase appears.
transition point. By
using the scaling in the RS phase, Eq.~(\ref{220929_17May20}), we can
see that $\lambda_-$ for $\eta^2\ll p \ll \eta$ behaves as
\begin{align}
 \lambda_- \sim c_1 (\eta^{-1}p)^2 -p,
\end{align}
where $c_1$ denotes a positive constant, meaning that the RSB occurs at
\begin{align}
 p_{\rm RSB}\sim \eta^2.
\end{align}
This is consistent with the numerical solution of the RS equations
presented in the main text.
\section{Full RSB Analysis}
\label{084946_15Jul20}

Here we calculate the contact number $z$ as a function of $p$
in the RSB phase by directly analyzing the full RSB free energy.

\subsection{Free energy}
For the most general form of Ansatz, $Q_{ab}$ is parameterized by a
continuous function $q(x)$, $x\in [0,1]$~\cite{franz2017universality}.
Let we assume that $q(x)$ is a continuous function for $x\in [x_m,
x_M]$.  In this interval, we can consider the inverse function $x(q)$.
Following the same procedure in Ref.~\cite{franz2017universality}, we
can write the free-energy as a functional of $x(q)$:
\begin{align}
 -\beta F[x(q)] &= \frac{1}{2}\left[\log(q_0-q_M)
 + \frac{q_m}{\lambda(q_m)}+\int_{q_m}^{q_M}\frac{dq}{\lambda(q)}\right]
 + \left.\alpha \gamma_{q_m+\eta^2}*f(q_m,h)\right|_{h=q_0/2-{a}}\new
 &-\alpha \int dh P(q_M,h)\left[f(q_M,h)-\log \gamma_{q_0-q_M}*e^{-\beta v(h)}\right]\new
 &+\alpha \int dh \int_{q_m}^{q_M}dq P(q,h)\left[\pdiff{f(q,h)}{q}
 + \frac{1}{2}\pdiff{^2 f(q,h)}{h^2}
 +\frac{x(q)}{2}\left(\pdiff{f(q,h)}{h}\right)^2\right],
\end{align}
where $q_m = q(x_m)$ and $q_M = q(x_M)$, and 
\begin{align}
 \lambda(q) = 1-q_M +\int_q^{q_M}dp x(p).
\end{align}
The functions $f(q,h)$ and $P(q,h)$ are determined by the so-called Parisi equations:
\begin{align}
 &\pdiff{f(q,h)}{q} = -\frac{1}{2}\left[\pdiff{^2f(q,h)}{h^2}+ x(q)\left(\pdiff{f(q,h)}{h}\right)^2\right], \new
 & \pdiff{P(q,h)}{q} = \frac{1}{2}\pdiff{}{h}\left[\pdiff{P(q,h)}{h}
 -2x(q)\left(P(q,h)\pdiff{f(q,h)}{h}\right)\right],\label{162220_16May20}
\end{align}
with the boundary conditions:
\begin{align}
& f(q_M,h) = \log \gamma_{q_0-q_M}*e^{-\beta v(h)},\new
& P(q_m,h) = \gamma_{q_m+\eta^2}(q_0/2-{a}-h).
\end{align}
The saddle point condition for $x(q)$ leads to 
\begin{align}
& \frac{q_m}{\lambda(q_m)^2} + \int_{q_m}^q \frac{dp}{\lambda(p)^2}
 = \alpha \int dh P(q,h)\left(\pdiff{f(q,h)}{h}\right)^2.\label{062431_16Mar20}
\end{align}
In the full RSB phase, $x(q)$ has a continuous part, which allows us to
calculate the derivative of Eq.~(\ref{062431_16Mar20})
w.r.t. $q$:
\begin{align}
 \frac{1}{\lambda(q)^2} = \alpha \pdiff{}{q}\int dh P(q,h)\left(\pdiff{f(q,h)}{h}\right)^2.\label{162301_16May20}
\end{align}
Using Eqs.~(\ref{162220_16May20}), after some manipulations,
Eq.~(\ref{162301_16May20}) can be rewritten as
\begin{align}
& \frac{1}{\lambda(q)^2} = \alpha \int dh P(q,h)\left(\pdiff{^2 f(q,h)}{h^2}\right)^2.\label{121026_16Mar20}
\end{align}
Contrarily, the saddle point condition for $q_0$ leads to
\begin{align}
 \frac{1}{q_0-q_M} = -\alpha \int dhP(q_M,h)\pdiff{f(q_M,h)}{h} -\alpha \int dh P(q_M,h)\pdiff{^2 f(q_M,h)}{h^2}.\label{120522_16Mar20}
\end{align}

\subsection{Zero temperature limit}

The equations can be further simplified in the zero temperature limit
$T\to 0$. We consider the harmonic expansion as in the case of the RS
analysis:
\begin{align}
 q_0-q_M \sim T\chi.\label{120457_16Mar20}
\end{align}
Then, we get 
\begin{align}
 &f(q_M,h) \sim -\frac{\beta h^2}{2(1+\chi)}\theta(-h),\\
 &g(h)  = \fdiff{F}{v(h)}\sim
 \begin{cases}
  \alpha (1+\chi)P(q_0,(1+\chi)h) & {\rm for}\ h<0\\
  \alpha P(q_0,h) & {\rm for}\ h>0.
 \end{cases}\label{120503_16Mar20}
\end{align}
Substituting Eqs.~(\ref{120457_16Mar20})--(\ref{120503_16Mar20}) into Eq.~(\ref{121026_16Mar20}), we get
\begin{align}
 \left(\frac{1+\chi}{\chi}\right)^2 = \int_{-\infty}^0 dhg(h) \to z = \left(1+\frac{1}{\chi}\right)^2.\label{121150_16Mar20}
\end{align}
Substituting Eqs.~(\ref{120457_16Mar20})--(\ref{120503_16Mar20}) into Eq.~(\ref{120522_16Mar20}), we get
\begin{align}
 &\frac{1+\chi}{\chi} = (1+\chi)\int_{-\infty}^0 dh g(h)h + \int_{-\infty}^0 dh g(h)\new
 &\to p = \frac{z}{1+\chi}-\frac{1}{\chi}.\label{121154_16Mar20}
\end{align}
From Eqs.(\ref{121150_16Mar20}) and (\ref{121154_16Mar20}), 
we get
\begin{align}
  z = \left(1+\sqrt{p}\right)^2.\label{112842_13May20}
\end{align}
Furthermore, by substituting Eq.~(\ref{112842_13May20}) into
$\lambda_-$, one can see that $\lambda_-=0$ in the RSB phase.

\end{document}